\documentclass{ws-procs9x6}

\begin{document}

\title{SEARCH FOR CP AND CPT VIOLATION IN POSITRONIUM DECAY}

\author{G.S.\ ADKINS$^*$}

\address{Department of Physics and Astronomy, Franklin \& Marshall College,\\
Lancaster, Pennsylvania 17604, USA\\
$^*$E-mail: gadkins@fandm.edu}

\begin{abstract}
Positronium, the electron-positron bound state, is described to a good approximation
by pure QED.  The states of positronium have definite values of C and P.  Consequently,
positronium is an attractive system for the investigation of possible violations of the
discrete symmetries in the leptonic sector.  We discuss signals for CP and CPT violation
in the decay of spin-polarized orthopositronium and show where such correlations might
arise in the context of the Standard-Model Extension.
\end{abstract}

\bodymatter

\section{Introduction}
Positronium has long been used for precision tests of QED and in searches for violations 
of the discrete symmetries C, P, and T, singly and in various combinations.  The physics 
of positronium is governed almost completely by QED, so that any sizable discrepancies 
between predictions and measured quantities would be signs of new physics.  The bound 
states of positronium have simple properties under P and C, which makes positronium a 
particularly attractive system for tests of the discrete symmetries.

Tests of the discrete symmetries in positronium have taken three basic forms:

\begin{enumerate}

\item Search for decays to the ÒwrongÕÕ number of photons.  (See Ref.\ \refcite{gsa:Gninenko06} 
for a recent review.)  Since positronium states are eigenstates of C, with eigenvalues $C=(-1)^{\ell+s}$, ground 
state spin-singlet positronium---parapositronium---has even parity under C, and decays to an even number of photons, while ground state spin-triplet positronium---orthopositronium (o-Ps)---has odd C parity, and decays to an odd number of photons.

\begin{figure}
\begin{center}
\psfig{file=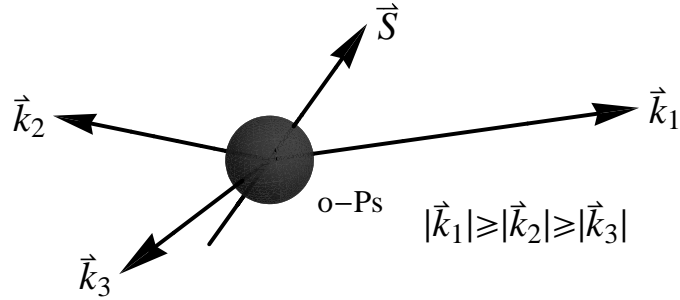,width=4.0in,bbllx=0bp,bblly=30bp,bburx=310bp,bbury=160bp,clip=}
\end{center}
\caption{Decay of orthopositronium with initial spin $\vec S$ into three photons..}
\label{gsa:fig1}
\end{figure}

\item Tests involving transition rates. \cite{gsa:Conti93}  These can involve searches for forbidden transitions or for asymmetries caused by symmetry violation.

\item Tests involving correlations of the momenta or polarizations of the final state photons for decays from spin-polarized orthopositronium (see Fig.\ \ref{gsa:fig1}).  Some measurable correlations are allowed, such as $\langle \hat k_i \cdot \hat k_j \rangle $ or $\langle \vec S \cdot \vec h_1 \rangle $ where $\hat k_i$ is a unit vector in the direction of the momentum of the $i^{th}$ final state photon, $\vec h_i$ describes the helicity of the $i^{th}$ photon, and $\vec S$ describes the orthopositronium spin.  (The photons are numbered in order of decreasing energy.)  QED predictions for the allowed correlations can be obtained using standard methods,\cite{gsa:Adkins10} and provide both an opportunity for precision tests of bound state QED and known quantities for testing techniques used to search for the forbidden correlations.  Correlations such as $\langle ( \hat S \cdot \hat k_1 ) ( \hat S \cdot \hat k_1 \times \hat k_2 ) \rangle$ and $\langle \vec S \cdot \hat k_1 \times \hat k_2 \rangle$ are forbidden \cite{gsa:Bernreuther88a,gsa:Bernreuther88b} and a non-vanishing value (apart from the effects of final state interactions) would be evidence for symmetry violation.

\end{enumerate}

Here we review the most recent (null) measurements of the symmetry violating correlations and discuss their dependence on the various Standard-Model Extension (SME)\cite{gsa:sme} coefficients.

\section{Symmetry violating correlations}
\label{gsa:sec2}

Two symmetry violating correlations have been extensively studied:

\subsection{$\langle \vec S \cdot \hat k_1 \times \hat k_2 \rangle$}

This correlation is CP symmetric but violates CPT.  A non-vanishing correlation implies a difference between the number of decays with the normal to the decay plane parallel ($+$) to the spin direction and the number with the normal antiparallel ($-$) to the spin direction.  The measured quantity is the asymmetry
\begin{equation}
A=\frac{N_+ - N_-}{N_+ + N_-} .
\end{equation}
The CPT violating coefficient was obtained from $A$ by including a factor representing the ``analyzing power'' of the experiment.  The asymmetry was measured by several groups starting with Arbic {\it et al.} in 1988.\cite{gsa:Arbic88}.  They found the CPT violating angular coefficient 
\begin{equation}
C_n = 0.020 \pm 0.023 .
\end{equation}
Andrukhovich, Antovich, and Berestov reported a measurement of the asymmetry in 2000: $A=0.0008 \pm 0.00091$.\cite{gsa:Andrukhovich00}  This was quoted by Vetter and Freedman \cite{gsa:Vetter03} as giving a coefficient of 
\begin{equation}
C_n = 0.0140 \pm 0.0190 .
\end{equation}
Finally, the 2003 result of Vetter and Freedman, \cite{gsa:Vetter03} measured using the Gammasphere, was
\begin{equation}
C_n = 0.0071 \pm 0.0062 .
\end{equation}

\subsection{$\langle ( \hat S \cdot \hat k_1 ) ( \hat S \cdot \hat k_1 \times \hat k_2 ) \rangle$}

This correlation violates P, T, and CP, but not CPT.  It was first measured by Skalsey and Van House in 1991.\cite{gsa:Skalsey91}  They found a CP violating coefficient of
\begin{equation}
C_{\rm CP} = -0.0056 \pm 0.0154 .
\end{equation}
The most recent measurement was by Yamazaki {\it et al.} in 2010.\cite{gsa:Yamazaki10}.  Their result was
\begin{equation}
C_{\rm CP} = 0.0013 \pm 0.0022 ,
\end{equation}
giving a factor of seven improvement.  Both results are consistent with zero.

\section{Analysis of the symmetry-violating correlations}

The Standard Model (SM) gives small but non-vanishing contributions for the correlations of Sec.\ \ref{gsa:sec2}.  The operator $\vec S \cdot \hat k_1 \times \hat k_2$ is odd under time reversal, but the o-Ps states are unstable and are not eigenstates of T.   The correlation $\langle \vec S \cdot \hat k_1 \times \hat k_2 \rangle$ has a contribution at the $O(\alpha^2)$ level arising from final-state interactions involving standard QED light-by-light scattering. \cite{gsa:Bernreuther88b}  The CP-violating correlation $\langle ( \hat S \cdot \hat k_1 ) ( \hat S \cdot \hat k_1 \times \hat k_2 ) \rangle$ has small non-vanishing weak interaction contributions. \cite{gsa:Bernreuther88b}

The Standard Model Extension provides a context for thinking about CPT and CP violation beyond the SM.  In distinction to SM effects, SME contributions would be expected to depend on sidereal time.  The discrete symmetry properties of the correlation operators of Sec.\ \ref{gsa:sec2} are shown in Table \ref{gsa:tab1}.  The SME operators that share these symmetry properties have coefficients $b_j$, $g_{j 0 \ell}$, $g_{j k 0}$, $(k_{AF})_j$ for $\vec S \cdot \hat k_1 \times \hat k_2$ and $c_{0 j}$, $c_{j 0}$, $(k_F)_{0 j k \ell}$ for $(\hat S \cdot \hat k_1)(\hat S \cdot \hat k_1 \times \hat k_2)$. \cite{gsa:Kostelecky02}  Explicit results for the symmetry violating correlations in the SME would be expected to involve these coefficients.  The corresponding calculations are underway.

\begin{table}
\tbl{Discrete symmetry properties of the correlation operators.}
{\begin{tabular}{@{}cccccc@{}}
\toprule
Operator & C & P & T & CP & CPT \\
\colrule
$\vec S \cdot \hat k_1 \times \hat k_2$ & $+$ & $+$ & $-$ & $+$ & $-$ \\
$(\hat S \cdot \hat k_1 ) (\hat S \cdot \hat k_1 \times \hat k_2 )$ & $+$ & $-$ & $-$ & $-$ & $+$ \\
\botrule
\end{tabular}}
\label{gsa:tab1}
\end{table}

\end{document}